\documentclass{article}

\usepackage{amsmath,amssymb}
\usepackage{graphicx}
\usepackage{color}
\usepackage[all]{xy}

\newtheorem{theorem}{Theorem}

\newtheorem{lemma}{Lemma}
\newtheorem{corollary}{Corollary}

\newtheorem{definition}{Definition}

\newtheorem{remark}{Remark}
\newtheorem{conjecture}{Conjecture}

\newcommand{\done}{\hfill $\Box$ }

\newcommand{\ls}[1]
    {\dimen0=\fontdimen6\the\font\lineskip=#1\dimen0
     \advance\lineskip.5\fontdimen5\the\font
     \advance\lineskip-\dimen0
     \lineskiplimit=0.9\lineskip
     \baselineskip=\lineskip
     \advance\baselineskip\dimen0
     \normallineskip\lineskip\normallineskiplimit\lineskiplimit
     \normalbaselineskip\baselineskip
     \ignorespaces}

\begin{document}

\bibliographystyle{abbrv}

\title{The Proof of Lin's Conjecture via the Decimation-Hadamard Transform}

\author{$^1$Honggang Hu, $^1$Shuai Shao, $^2$Guang Gong, and $^3$Tor Helleseth\\
\and
$^1$School of Information Science and Technology\\
University of Science and Technology of China\\
Hefei, China, 230027\\
Email. hghu2005@ustc.edu.cn\\
\and
$^2$Department of Electrical and Computer Engineering \\
University of Waterloo \\
Waterloo, Ontario N2L 3G1, Canada \\
Email. ggong@uwaterloo.ca\\
\and
$^3$The Selmer Center
\\Department of Informatics\\
University of Bergen\\
PB 7803, N-5020 Bergen, Norway\\
Email. Tor.Helleseth@ii.uib.no}

\date{}
 \maketitle

\thispagestyle{plain}
\setcounter{page}{1}

\begin{abstract}
 In 1998, Lin presented a conjecture on a class of ternary sequences with ideal 2-level autocorrelation in his Ph.D thesis. Those sequences have a very simple structure, i.e., their trace representation has two trace monomial terms. In this paper, we present a proof for the conjecture. The mathematical tools employed are the second-order multiplexing decimation-Hadamard transform, Stickelberger's theorem,  the Teichm\"{u}ller character, and  combinatorial techniques for enumerating the Hamming weights of ternary numbers.  As a by-product, we also prove that the Lin conjectured ternary sequences are Hadamard equivalent to ternary $m$-sequences.
\end{abstract}


\ls{1.5}
\section{Introduction}\label{sec_intro}

Sequences with good random properties have wide applications in modern communications and cryptography, such as CDMA communication systems, global positioning systems, radar, and stream cipher cryptosystems \cite{Golomb82, GG2005, SOSL85}. The research of new sequences with good correlation properties has been an interesting research issue for decades, especially sequences with ideal two-level autocorrelation \cite{GG2005, HK98}.

There has been significant progress in finding new sequences with ideal two-level autocorrelation in the last two decades. In 1997, by exhaustive search, Gong, Gaal and Golomb found a class of binary sequences of period $2^n-1$ with 2-level autocorrelation in \cite{ggg97}, and in 1998, No, Golomb, Gong, Lee, and Gaal published  five conjectures regarding binary sequences of period $2^n-1$ with ideal two-level
autocorrelation \cite{NGGLG98} including  two classes, called {\em Welch-Gong transformation sequences}, conjectured by the group of the authors in \cite{ggg97}.   Interestingly, using monomial
hyperovals, Maschietti constructed three classes of  binary sequences of period $2^n-1$ with ideal two-level autocorrelation \cite{Maschietti98} from Segre and Green type monomial hyper ovals and a shorter proof of those sequences is reported in \cite{Xiang98} \cite{Dillon99}.  Shortly after that, No, Chung, and Yun  \cite{NCY98}, in terms of  the image set of the  polynomial $z^d+(z+1)^d$ where $d=2^{2k}-2^k+1$  where $3k\equiv 1 \bmod{n}$, a special Kasami exponent,
conjectured another class of binary sequences of period $2^n-1$ with ideal two-level autocorrelation. This class  turned out to be the same class as  the Welch-Gong sequences conjectured in \cite{NGGLG98} and Dobbertin formally proved that in \cite{Dobbertin992}.
In 1999, for the case of $n$ odd, Dillon proved the conjecture of Welch-Gong sequence using the Hadamard transform \cite{Dillon99}, i.e., he showed that the Welch-Gong sequence is equivalent to an $m$-sequence under the Hadamard transform. A few months later, Dillon and Dobbertin confirmed all these conjectured classes of ideal two-level autocorrelation sequences of period $2^n-1$, although the paper is published later \cite{DD04}.  The progress on binary 2-level autocorrelation sequences has been collected in \cite{GG2005} and has no new sequences coming out since then.

The progress on searching for  nonbinary sequences with 2-level autocorrelation seems different.  For $p=3$, Lin conjectured a class of ideal two-level autocorrelation sequences of period $3^n-1$ with two trace monomial terms  in 1998 in his Ph.D thesis \cite{Lin98}. In 2001, a new class of ternary ideal two-level autocorrelation sequences of period $3^n-1$ was constructed by Helleseth, Kumar, and Martinsen \cite{HKM01}. In 2001, Ludkovski and Gong proposed several conjectures regarding ternary sequences with ideal two-level autocorrelation \cite{LG01},  which are obtained by applying the second order decimation and Hadamard transform, introduced in \cite{GG02}.

For any $p \ne 2$, in \cite{HG02}, Helleseth and Gong found a construction of $p$-ary sequences of period $p^n-1$ with ideal two-level autocorrelation which includes the construction in \cite{HKM01} when $p=3$. For the ternary case, the validity of  the Lin conjectured sequences has been first announced by Dillon, Arasu and Player in 2004 \cite{ArasuDillonPlayer2004}.
   Together with Lin's conjecture, those found by Ludkovski and Gong  have been claimed   recently by Arasu in  \cite{Arasu11} for  which it is referred to an unpublished paper by Arasu, Dillon and Player  \cite{ArasuDillonPlayer}.  Nevertheless,  the proofs have not appeared in the public domain yet  since  SETA 2004 announced this result \cite{ArasuDillonPlayer2004} in 2004.  Their approach is to use  the Gauss sum and group ring to represent sequences as many researchers do, say \cite{Xiang99}, to just  list a few, and  the Hasse-Davenport identity  to determine the trace representation of the sequences.

In this paper, we provide a proof for the  Lin conjecture  through the decimation Hadamard transform.  In 2002, Gong and Golomb introduced the concept of  the iterative decimation-Hadamard
transform (DHT) to investigate ideal two-level autocorrelation sequences \cite{GG02}. They showed that, for all odd $n\leq 17$,
using the second-order DHT and starting with a single binary $m$-sequence, one can obtain all known binary ideal two-level autocorrelation sequences of period $2^n-1$ without subfield factorization. Later, Yu and Gong generalized the second-order DHT to the second-order multiplexing DHT \cite{YG05, YG08}. In this paper, we prove that, using the second-order multiplexing DHT and starting with a single ternary $m$-sequence, one may obtain the Lin conjectured ternary ideal two-level autocorrelation sequences.  The second set of the  key tools for the proof are Stickelberger's theorem and the Teichm\"{u}ller character. Elementary enumeration methods for ternary numbers play the essential role in the last touch of  the proof.
Those methods  are   different  from the approach sketched  in \cite{ArasuDillonPlayer2004}.  As a by-product,  we  also confirm Conjecture 2 in \cite{Gong2012} which is selected from \cite{five11}. In other words, the Lin conjectured ternary sequences are Hadamard equivalent to ternary $m$-sequences.

This paper is organized as follows. In Section \ref{sec_pre}, we give some
notation and background which will be used later. In Sections \ref{sec_lin_1} and \ref{sec_lin_2}, we
present the proof of the Lin  Conjecture. Finally, Section \ref{sec_con} concludes this
paper.

\section{Preliminaries}\label{sec_pre}

Let $\mathbb{F}_q$ denote the finite field of order $q$, where
$q=p^n$, and $p$ is a prime number, and $Tr(\cdot)$ denote the trace
map from $\mathbb{F}_q$ to $\mathbb{F}_p$. The primitive $p$th root of unity in characteristic 0 is
denoted as $\omega_p$, i.e., $\omega_p=e^{2\pi i/p}$.

\subsection{Ideal Two-Level Autocorrelation Sequence and Lin's Conjecture}

Let $S=\{s_i\}$ be an $p$-ary sequence with period $N$. For any $0\leq\tau<N$, the autocorrelation of $S$ at shift $\tau$ is defined by
$$C_S(\tau)=\sum_{i=0}^{N-1}\omega_p^{s_{i+\tau}-s_i}.$$
If $C_S(\tau)=-1$ for any $0<\tau<N$, we call $S$ an {\em (ideal) two-level autocorrelation sequence}.

\begin{conjecture}[Lin's Conjecture \cite{Lin98}]
Let $n=2m+1$, and $\alpha$ be a primitive element in $\mathbb{F}_{3^n}$. Suppose that $S=\{s_i\}$ is a ternary sequence defined by $s_i=Tr(\alpha^i+\alpha^{(2\cdot 3^m+1)i})$ for $i=0, 1, 2, \cdots$. Then $S$ has  ideal two-level autocorrelation.
\end{conjecture}

\subsection{The Second-Order Decimation-Hadamard Transform}

Let $f(x)$ be a polynomial from $\mathbb{F}_q$ to $\mathbb{F}_p$.
Then the Hadamard transform of $f(x)$ is defined by
$$\widehat{f}(\lambda)=\sum_{x\in \mathbb{F}_q}\omega_p^{Tr(\lambda x)-f(x)},\lambda\in \mathbb{F}_q,$$
and the inverse transform is given by
$$\omega_p^{f(\lambda)}=\frac{1}{q}\sum_{x\in \mathbb{F}_q}\omega_p^{Tr(\lambda x)}\overline{\widehat{f}(x)},\lambda\in \mathbb{F}_q.$$
The following three concepts are from \cite{GG02}.

\begin{definition}
For any integer $0<v<q-1$, we define
$$\widehat{f}(v)(\lambda)=\sum_{x\in \mathbb{F}_q}\omega_p^{Tr(\lambda x)-f(x^v)},\lambda\in \mathbb{F}_q.$$
$\widehat{f}(v)(\lambda)$ is called the {\em first-order
decimation-Hadamard transform (DHT)} of $f(x)$ with respect to
$Tr(x)$, and  the first-order DHT for short.
\end{definition}

\begin{definition}\label{def_second}
For any integers $0<v,t<q-1$, we define
$$\widehat{f}(v,t)(\lambda)=\sum_{y\in \mathbb{F}_q}\omega_p^{Tr(\lambda y)}\overline{\widehat{f}(v)(y^t)},\lambda\in \mathbb{F}_q,$$
where $\overline{\widehat{f}(v)(y^t)}$ is the complex conjugate of
$\widehat{f}(v)(y^t)$. $\widehat{f}(v,t)(\lambda)$ is called the {\em
second-order decimation-Hadamard transform (DHT)} of $f(x)$ with
respect to $Tr(x)$, and  the second-order DHT for short.
\end{definition}

\begin{remark}
If $t=1$, then $\widehat{f}(v,t)(\lambda)/q$ is just the inverse Hadamard transform
of $f(x)$.
\end{remark}

\begin{definition}
With the notation as in Definition \ref{def_second}, if
$$\widehat{f}(v,t)(\lambda)\in \{q\omega_p^i\ |\ i=0,1,\cdots  ,p-1\}, \lambda\in \mathbb{F}_q,$$
then $(v,t)$ is called a {\em realizable pair} of $f(x)$. In this case,
let
$$\omega_p^{g(x)}=\frac{1}{q}\widehat{f}(v,t)(x),x\in \mathbb{F}_q.$$
Then $g(x)$ is called a {\em realization} of $f(x)$ under $(v,t)$.
\end{definition}

\subsection{The Second-Order Multiplexing Decimation-Hadamard Transform}

For the case of $\gcd(v,q-1)>1$, we may define another
kind of decimation-Hadamard transform, namely, the multiplexing
decimation-Hadamard transform, which introduced are introduced in \cite{YG05, YG08}.

\begin{definition}
For any integer $0<v<q-1$ and $\gamma\in \mathbb{F}_{q}^{*}$, we
define
$$\widehat{f}(v)(\lambda, \gamma)=\sum_{x\in \mathbb{F}_q}\omega_p^{Tr(\lambda x)-f(\gamma x^v)},\lambda\in \mathbb{F}_q.$$
$\widehat{f}(v)(\lambda, \gamma)$ is called the {\em first-order
multiplexing decimation-Hadamard transform (DHT)} of $f(x)$ with
respect to $Tr(x)$, and the first-order multiplexing DHT
for short.
\end{definition}

\begin{definition}\label{def_mDHT_2}
For any integers $0<v,t<q-1$ and $\gamma\in \mathbb{F}_{q}^{*}$,
we define
$$\widehat{f}(v,t)(\lambda, \gamma)=\sum_{y\in \mathbb{F}_q}\omega_p^{Tr(\lambda y)}\overline{\widehat{f}(v)(y^t, \gamma)},\lambda\in \mathbb{F}_q,$$
where $\overline{\widehat{f}(v)(y^t, \gamma)}$ is the complex
conjugate of $\widehat{f}(v)(y^t, \gamma)$.
$\widehat{f}(v,t)(\lambda, \gamma)$ is called the {\em second-order
multiplexing decimation-Hadamard transform (DHT)} of $f(x)$ with
respect to $Tr(x)$, and the second-order multiplexing DHT
for short.
\end{definition}

\begin{definition}
With the notation as in Definition \ref{def_mDHT_2}, if
$$\widehat{f}(v,t)(\lambda, \gamma)\in \{q\omega_p^i\ |\ i=0,1,\cdots  ,p-1\}, \lambda\in \mathbb{F}_q, \gamma\in \mathbb{F}_q^{*}$$
then $(v,t)$ is called a realizable pair of $f(x)$. In this case,
let
$$\omega_p^{g(x, \gamma)}=\frac{1}{q}\widehat{f}(v,t)(x, \gamma),x\in \mathbb{F}_q.$$
Then $g(x, \gamma)$ is called a {\em realization} of $f(x)$ under $(v,t)$
and $\gamma$.
\end{definition}

\subsection{Gauss Sums and Stickelberger's Theorem}

The mapping $\psi$ defined by
$$\psi(x)=\omega_p^{Tr(x)}$$
is an additive character of $\mathbb{F}_q$. Suppose that $\chi$ is a
multiplicative character of $\mathbb{F}_q^{*}$. For the convenience, we extend $\chi$ to
$\mathbb{F}_q$ by defining $\chi(0)=0$. Henceforth, the multiplicative character set of $\mathbb{F}_q^{*}$ will be denoted by $\widehat{\mathbb{F}_q^{*}}$ for simplicity.

\begin{definition}
For any multiplicative character $\chi$ over $\mathbb{F}_q$, the
Gauss sum $G(\chi)$ over $\mathbb{F}_q$ is defined by
$$G(\chi)=\sum_{x\in F_q}\psi(x)\chi(x).$$
\end{definition}

\begin{lemma}[\cite{LN83}]\label{lem_gauss}
For any multiplicative character $\chi$ over $\mathbb{F}_q$, we have
$$G(\overline{\chi})=\chi(-1)\overline{G(\chi)}\mbox{ and }G(\chi^p)=G(\chi).$$
If $\chi$ is trivial, then $G(\chi)=-1$. Furthermore, if $\chi$ is
nontrivial, then
$$G(\chi)\overline{G(\chi)}=q.$$
In other words, for any nontrivial character $\chi$, $G(\chi)$ is
invertible, and $G(\chi)^{-1}=\overline{G(\chi)}/q$.
\end{lemma}

\begin{figure}
\centering
$$\xymatrix{
& \mathbb{Z}[\omega_p, \omega_{q-1}] & \\
  \mathbb{Z}[\omega_p] \ar[ur]^{(\pi)=\mathcal{Q}_1\mathcal{Q}_2\cdots  \mathcal{Q}_t}  &  &    \mathbb{Z}[\omega_{q-1}] \ar[ul]_{\mathfrak{p}_i=\mathcal{Q}_i^{p-1}}    \\
                & \ar[ul]^{(p)=\pi^{p-1}} \mathbb{Z}    \ar[ur]_{(p)=\mathfrak{p}_1\mathfrak{p}_2\cdots  \mathfrak{p}_t}
                }$$
\caption{Prime Ideal Factorization}\label{fig_primeideal}
\end{figure}
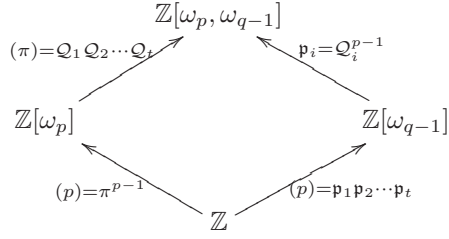

The factorization of prime ideals in algebraic
integer rings is an interesting issue. $(p)$ is a prime ideal in $\mathbb{Z}$. Let
$\pi=\omega_p-1$. It is known that $(\pi)$ is a prime ideal in
$\mathbb{Z}[\omega_p]$. Moreover, $(p)=(\pi)^{p-1}$ in
$\mathbb{Z}[\omega_p]$, and
$(\pi)=\mathcal{Q}_1\mathcal{Q}_2\cdots  \mathcal{Q}_t$ in
$\mathbb{Z}[\omega_p, \omega_{q-1}]$, where $\mathcal{Q}_i$ are
prime ideals in $\mathbb{Z}[\omega_p, \omega_{q-1}]$, and
$t=\phi(p^n-1)/n$. Hence,
$(p)=(\mathcal{Q}_1\mathcal{Q}_2\cdots  \mathcal{Q}_t)^{p-1}$ in
$\mathbb{Z}[\omega_p, \omega_{q-1}]$. On the other hand,
$(p)=\mathfrak{p}_1\mathfrak{p}_2\cdots  \mathfrak{p}_t$ in
$\mathbb{Z}[\omega_{q-1}]$. For each $\mathfrak{p}_i$, it is the
$(p-1)$-th power of a prime ideal in $\mathbb{Z}[\omega_p,
\omega_{q-1}]$. Without loss of generality, we may assume that
$\mathfrak{p}_i=\mathcal{Q}_i^{p-1}$. For the relationship among $(p),
\mathfrak{p}_i$, and $\mathcal{Q}_i$, the reader is referred to Figure
\ref{fig_primeideal}.

For each $\mathcal{Q}_i$, we have $\mathbb{Z}[\omega_p,
\omega_{q-1}]/\mathcal{Q}_i\cong \mathbb{F}_q$ because
$[\mathbb{Z}[\omega_p, \omega_{q-1}]/\mathcal{Q}_i:
\mathbb{Z}/(p)]=n$. Henceforth, we fix one prime ideal
$\mathcal{Q}_i$, and denote it by $\mathcal{Q}$ for simplicity.
There is one special multiplicative character $\chi$ on $\mathbb{F}_q$
satisfying
$$\chi(x)(\mbox{mod }\mathcal{Q})=x.$$
This character is called the {\em Teichm\"{u}ller character}. For simplicity, henceforth we
denote it by $\chi_\mathfrak{p}$. The Teichm\"{u}ller character has been used to investigate the dual of certain bent functions \cite{GHHK12}.

For any $0\leq k<q-1$, let $k=k_0+k_1p+\cdots+k_{n-1}p^{n-1}$ be the
$p$-adic representation of $k$, where $0\leq k_i<p$ for
$i=0,1,\dots,n-1$. Let $\mathrm{wt}(k)=k_0+k_1+\cdots+k_{n-1}$, and $\sigma(k)=k_0!k_1!\cdots k_{n-1}!$. Moreover, for any $j$, we use
$\mathrm{wt}(j)$ and $\sigma(j)$ to denote
$\mathrm{wt}(\overline{j})$ and $\sigma(\overline{j})$ respectively, where $0\leq \overline{j}<q-1$
and $j\equiv \overline{j}\ (\bmod\;q-1)$.

\begin{theorem}[Stickelberger's Theorem, \cite{Lang78}]
For any $0<k<q-1$, we have
$$G(\chi_\mathfrak{p}^{-k})\equiv -\frac{\pi^{wt(k)}}{\sigma(k)}(\mbox{mod }\pi^{wt(k)+p-1}).$$
\end{theorem}

Let $e=\lfloor wt(k)/(p-1)\rfloor$, where $\lfloor\cdot\rfloor$ is
the floor function. Then $p^e\|G(\chi_\mathfrak{p}^{-k})$ for any
$0<k<q-1$ by Stickelberger's theorem. The following lemma is extremely powerful, which will be used later.

\begin{lemma}[\cite{HHKWX09}]\label{lem_gauss_trace}
For any $y\in \mathbb{F}_{q}^{*}$, we have
$$\omega_p^{Tr(y)}=\frac{1}{q-1}\sum_{\chi\in\widehat{\mathbb{F}_{q}^{*}}}G(\chi)\overline{\chi}(y).$$
\end{lemma}

\section{Proof of the Lin Conjecture: Part I}\label{sec_lin_1}

\begin{lemma}\label{lem_ring}
For any $e\in \mathbb{Z}[\omega_3]$, if $3^n|e$, then $e=0$ or
$|e|\geq 3^n.$
\end{lemma}

{\bf Proof. } If $e\neq 0$, then $e=3^nf$ with $f\in
\mathbb{Z}[\omega_3]$ and $f\neq 0$. Let $f=f_0+f_1\omega_3$, where
$f_0, f_1\in \mathbb{Z}$. Then $|f|^2=f_0^2+f_1^2+f_0f_1\geq 1$.
Thus, $|e|=3^n|f|\geq 3^n$. \done

\begin{lemma} \label{lem_sum}
If $\gcd(t, 3^n-1)=1$, then for any $\gamma\in \mathbb{F}_{3^n}^{*}$
$$\sum_{\lambda\in \mathbb{F}_{3^n}}|\widehat{f}(v,t)(\lambda, \gamma)|^2=3^{3n}.$$
\end{lemma}

{\bf Proof. }
\begin{eqnarray*}
\sum_{\lambda\in
\mathbb{F}_{3^n}}|\widehat{f}(v,t)(\lambda, \gamma)|^2&=&\sum_{\lambda\in
\mathbb{F}_{3^n}}\sum_{x_1, y_1\in
\mathbb{F}_{3^n}}\omega_3^{Tr(\lambda y_1-y_1^t
x_1+\gamma x_1^v)}\sum_{x_1, y_2\in
\mathbb{F}_{3^n}}\omega_3^{Tr(-\lambda y_2+y_2^t x_2-\gamma x_2^v)}\\
&=&\sum_{x_1, x_2, y_1, y_2\in \mathbb{F}_{3^n}}\omega_3^{Tr(-y_1^t
x_1+\gamma x_1^v+y_2^t x_2-\gamma x_2^v)}\sum_{\lambda\in
\mathbb{F}_{3^n}}\omega_3^{Tr(\lambda y_1-\lambda y_2)}\\
&=&3^n\sum_{x_1, x_2, y\in \mathbb{F}_{3^n}}\omega_3^{Tr(-y^t
x_1+y^t x_2+\gamma x_1^v-\gamma x_2^v)}\\
&=&3^n\sum_{x_1, x_2\in \mathbb{F}_{3^n}}\omega_3^{Tr(\gamma x_1^v-\gamma x_2^v)}\sum_{y\in \mathbb{F}_{3^n}}\omega_3^{Tr(-y^t
x_1+y^t x_2)}\\
&=&3^{2n}\sum_{x_1, x_2\in \mathbb{F}_{3^n}, x_1=x_2}\omega_3^{Tr(\gamma x_1^v-\gamma x_2^v)}=3^{3n}.
\end{eqnarray*}
\done

\begin{lemma}\label{lem_gauss-v}
If $d=\gcd(v, 3^n-1)>1$, then for any $\gamma\in \mathbb{F}_{3^n}^{*}$, we have
$$\sum_{x\in \mathbb{F}_{3^n}^{*}}\omega_3^{Tr(\gamma
x^v)}=\sum_{\chi\in\widehat{\mathbb{F}_{3^n}^{*}}, \chi^d=1}G(\chi)\overline{\chi}(\gamma).$$
\end{lemma}

{\bf Proof. }Firstly, we have
\begin{eqnarray*}
\sum_{x\in \mathbb{F}_{3^n}^{*}}\omega_3^{Tr(\gamma
x^v)}&=&\sum_{x\in \mathbb{F}_{3^n}^{*}}\omega_3^{Tr(\gamma
x^d)}.
\end{eqnarray*}
By Lemma \ref{lem_gauss_trace}, it follows that
\begin{eqnarray*}
\sum_{x\in \mathbb{F}_{3^n}^{*}}\omega_3^{Tr(\gamma
x^v)}&=&\sum_{x\in \mathbb{F}_{3^n}^{*}}\frac{1}{3^n-1}\sum_{\chi\in\widehat{\mathbb{F}_{3^n}^{*}}}G(\chi)\overline{\chi}(\gamma
x^d)\\
&=&\frac{1}{3^n-1}\sum_{\chi\in\widehat{\mathbb{F}_{3^n}^{*}}}G(\chi)\overline{\chi}(\gamma)\sum_{x\in \mathbb{F}_{3^n}^{*}}\overline{\chi}^d(x)\\
&=&\sum_{\chi\in\widehat{\mathbb{F}_{3^n}^{*}}, \chi^d=1}G(\chi)\overline{\chi}(\gamma).
\end{eqnarray*}
\done

\begin{theorem}\label{thm_general}
Let $f(x)=Tr(x)$. For the multiplexing DHT of $f(x)$, if $\gcd(v, 3^n-1)>1$ and $\gcd(t,
3^n-1)=1$, then $(v, t)$ is a realizable pair if and only if
$wt(jvt)+wt(-jv)+wt(j)>2n$ for any $0<j<3^n-1$ with $jd\neq 0$,
where $d=\gcd(v, 3^n-1)$. Moreover, for any $\gamma\in \mathbb{F}_{3^n}^{*}$, the realization of $f(x)$ under $(v,t)$
and $\gamma$ is
given by $$g(v,t)(\lambda, \gamma)=\sum_{\begin{array}{c}
\ wt(jvt)+wt(-jv)+wt(j)\\
=2n+1, 0<j<3^n-1
\end{array}}(-1)^{jv}\sigma(jvt)\sigma(-jv)\sigma(j)(\gamma\lambda^{vt})^j.$$
\end{theorem}

{\bf Proof. }If $\lambda=0$, then $\widehat{f}(v,t)(\lambda, \gamma)=3^n$. For any $\lambda\neq 0$,
$$\widehat{f}(v,t)(\lambda, \gamma)=\sum_{x, y\in \mathbb{F}_{3^n}}\omega_3^{Tr(\lambda y)-Tr(y^tx)+Tr(\gamma x^v)}.$$
By Lemma \ref{lem_gauss_trace}, we have the following deviations.
\begin{eqnarray}
\widehat{f}(v,t)(\lambda, \gamma)&=&\sum_{x\in
\mathbb{F}_{3^n}^{*}}\sum_{y\in
\mathbb{F}_{3^n}}\omega_3^{Tr(\lambda y)-Tr(y^tx)+Tr(\gamma x^v)}\nonumber\\
&=&\sum_{x\in \mathbb{F}_{3^n}^{*}}\omega_3^{Tr(x^v)}+\sum_{x\in
\mathbb{F}_{3^n}^{*}}\sum_{y\in
\mathbb{F}_{3^n}^{*}}\omega_3^{Tr(\lambda y)-Tr(y^tx)+Tr(\gamma x^v)}\nonumber\\
&=&\sum_{x\in \mathbb{F}_{3^n}^{*}}\omega_3^{Tr(\gamma
x^v)}+\frac{1}{(3^n-1)^3}\sum_{x\in \mathbb{F}_{3^n}^{*}}\sum_{y\in
\mathbb{F}_{3^n}^{*}}\sum_{\chi_1\in\widehat{\mathbb{F}_{3^n}^{*}}}G(\chi_1)\overline{\chi_1}(\lambda
y)\sum_{\chi_2\in\widehat{\mathbb{F}_{3^n}^{*}}}\overline{G(\chi_2)}\chi_2(y^tx)\sum_{\chi_3\in\widehat{\mathbb{F}_{3^n}^{*}}}G(\chi_3)\overline{\chi_3}(\gamma x^v)\nonumber\\
&=&\sum_{x\in \mathbb{F}_{3^n}^{*}}\omega_3^{Tr(\gamma
x^v)}+\frac{1}{(3^n-1)^3}\sum_{\chi_1, \chi_2,
\chi_3\in\widehat{\mathbb{F}_{3^n}^{*}}}G(\chi_1)\overline{G(\chi_2)}G(\chi_3)\sum_{x\in
\mathbb{F}_{3^n}^{*}}\sum_{y\in
\mathbb{F}_{3^n}^{*}}\overline{\chi_1}(\lambda
y)\chi_2(y^tx)\overline{\chi_3}(\gamma x^v)\nonumber\\
&=&\sum_{x\in \mathbb{F}_{3^n}^{*}}\omega_3^{Tr(\gamma
x^v)}+\frac{1}{(3^n-1)^2}\sum_{\chi_1,
\chi_3\in\widehat{\mathbb{F}_{3^n}^{*}}}G(\chi_1)\overline{G(\chi_3^v)}G(\chi_3)\sum_{y\in
\mathbb{F}_{3^n}^{*}}\overline{\chi_1}(\lambda y)\chi_3(y^{vt})\overline{\chi_3}(\gamma)\nonumber\\
&=&\sum_{x\in \mathbb{F}_{3^n}^{*}}\omega_3^{Tr(\gamma x^v)}+\frac{1}{3^n-1}\sum_{\chi\in\widehat{\mathbb{F}_{3^n}^{*}}}G(\chi^{vt})\overline{G(\chi^v)}G(\chi)\overline{\chi}^{vt}(\lambda)\overline{\chi}(\gamma)\nonumber\\
&=&\sum_{x\in \mathbb{F}_{3^n}^{*}}\omega_3^{Tr(\gamma
x^v)}+\frac{1}{3^n-1}\sum_{\chi^d=1}G(\chi)\overline{\chi}(\gamma)+\frac{1}{3^n-1}\sum_{\chi^d\neq
1}G(\chi^{vt})\overline{G(\chi^v)}G(\chi)\overline{\chi}^{vt}(\lambda)\overline{\chi}(\gamma)\nonumber.
\end{eqnarray}
According to Lemma \ref{lem_gauss-v}, it follows that
\begin{eqnarray}
\widehat{f}(v,t)(\lambda, \gamma)
&=&\sum_{x\in \mathbb{F}_{3^n}^{*}}\omega_3^{Tr(\gamma
x^v)}+\frac{1}{3^n-1}\sum_{x\in \mathbb{F}_{3^n}^{*}}\omega_3^{Tr(\gamma
x^v)}+\frac{1}{3^n-1}\sum_{\chi^d\neq
1}G(\chi^{vt})\overline{G(\chi^v)}G(\chi)\overline{\chi}^{vt}(\lambda)\overline{\chi}(\gamma)\nonumber\\
&=&\frac{3^n}{3^n-1}\sum_{x\in \mathbb{F}_{3^n}^{*}}\omega_3^{Tr(\gamma
x^v)}+\frac{1}{3^n-1}\sum_{\chi^d\neq
1}G(\chi^{vt})\overline{G(\chi^v)}G(\chi)\overline{\chi}^{vt}(\lambda)\overline{\chi}(\gamma)\nonumber\\
&=&\frac{3^n}{3^n-1}\sum_{x\in \mathbb{F}_{3^n}^{*}}\omega_3^{Tr(\gamma
x^v)}+\frac{1}{3^n-1}\sum_{\chi^d\neq
1}G(\chi^{vt})G(\overline{\chi}^v)G(\chi)\overline{\chi}^{vt}(\lambda)\overline{\chi}(\gamma)\overline{\chi}^v(-1). \label{eqn_mDHT_ternary}
\end{eqnarray}
If $(v, t)$ is a realizable pair, then $\widehat{f}(v,t)(\lambda,
\gamma)(\mbox{mod }3^n)=0$ for any $\lambda$ and $\gamma\neq 0$. Thus, we have
$$\sum_{jd\neq
0}G(\overline{\chi}_\mathfrak{p}^{jvt})G(\chi_\mathfrak{p}^{jv})G(\overline{\chi}_\mathfrak{p}^{j})\chi_\mathfrak{p}^{jvt}(\lambda
)\chi_\mathfrak{p}^{j}(\gamma)\chi_\mathfrak{p}^{jv}(-1)(\mbox{mod
}3^n)=0$$ for any $\lambda\neq 0$ and $\gamma\neq 0$. Therefore,
$G(\overline{\chi}_\mathfrak{p}^{jvt})G(\chi_\mathfrak{p}^{jv})G(\overline{\chi}_\mathfrak{p}^{j})\chi_\mathfrak{p}^{jvt}(\lambda
)\chi_\mathfrak{p}^{jv}(-1)(\mbox{mod
}3^n)=0$ for any $jd\neq 0$ which is equivalent to
$wt(jvt)+wt(-jv)+wt(j)>2n$ for any $0<j<3^n-1$ with $jd\neq 0$.

On the other hand, if $wt(jvt)+wt(-jv)+wt(j)>2n$ for any $0<j<3^n-1$
with $jd\neq 0$, then $\widehat{f}(v,t)(\lambda, \gamma)(\mbox{mod
}3^n)=0$ for any $\lambda\neq 0$. Furthermore,
$(3^n-1)\widehat{f}(v,t)(\lambda, \gamma)(\mbox{mod
}\pi^{2n+1})=-3^n$ for any $\lambda\neq 0$. Thus,
$\widehat{f}(v,t)(\lambda, \gamma)\neq 0$ for any $\lambda\neq 0$. By Lemma \ref{lem_ring}, $|\widehat{f}(v,t)(\lambda, \gamma)|\geq3^n$. In addition, by Lemma \ref{lem_sum}, $\sum_{\lambda\in \mathbb{F}_{3^n}}|\widehat{f}(v,t)(\lambda, \gamma)|^2=3^{3n}$. Thus,
$|\widehat{f}(v,t)(\lambda, \gamma)|=3^n$ for any $\lambda$ and
$\gamma\neq 0$. Moreover, by (\ref{eqn_mDHT_ternary}),
$\widehat{f}(v,t)(\lambda)/3^n(\mbox{mod }\pi)=1$. Therefore, we
have $\widehat{f}(v,t)(\lambda)=3^n, 3^n\omega_3, 3^n\omega_3^2$
which means that $(v, t)$ is a realizable pair.

Let $\widehat{f}(v,t)(\lambda, \gamma)=3^n\omega_3^{g(v,t)(\lambda,
\gamma)}$, where $g(v,t)(\lambda, \gamma)=0, 1, 2$. Then
$\widehat{f}(v,t)(\lambda, \gamma)=3^n+3^ng(v,t)(\lambda,
\gamma)\pi+O(\pi^{2n+2})$, and $(3^n-1)f(v,t)(\lambda,
\gamma)=-3^n-3^{n}g(v,t)(\lambda, \gamma)\pi+O(\pi^{2n+2})$. By
(\ref{eqn_mDHT_ternary}),
\begin{eqnarray*}
(3^n-1)f(v,t)(\lambda, \gamma)&=&3^n\sum_{x\in \mathbb{F}_{3^n}^{*}}\omega_3^{Tr(\gamma
x^v)}+\sum_{\chi^d\neq
1}G(\chi^{vt})G(\overline{\chi}^v)G(\chi)\overline{\chi}^{vt}(\lambda)\overline{\chi}(\gamma)\overline{\chi}^v(-1)\\
&=&3^n\sum_{x\in \mathbb{F}_{3^n}^{*}}\omega_3^{Tr(\gamma
x^v)}+\sum_{jd\neq 0}G(\overline{\chi}_\mathfrak{p}^{jvt})G(\chi_\mathfrak{p}^{jv})G(\overline{\chi}_\mathfrak{p}^j)\chi_\mathfrak{p}^{jvt}(\lambda)\chi_\mathfrak{p}^j(\gamma)\chi_\mathfrak{p}^{jv}(-1).
\end{eqnarray*}
It follows that
\begin{eqnarray*}
g(v,t)(\lambda, \gamma)&=&\frac{(3^n-1)f(v,t)(\lambda, \gamma)+3^n}{-3^n\pi}(\mbox{mod }\mathcal{Q})\\
&=&\frac{3^n+3^n\sum_{x\in \mathbb{F}_{3^n}^{*}}\omega_3^{Tr(\gamma
x^v)}+\sum_{jd\neq 0}G(\overline{\chi}_\mathfrak{p}^{jvt})G(\chi_\mathfrak{p}^{jv})G(\overline{\chi}_\mathfrak{p}^j)\chi_\mathfrak{p}^{jvt}(\lambda)\chi_\mathfrak{p}^j(\gamma)\chi_\mathfrak{p}^{jv}(-1)}{-3^n\pi}(\mbox{mod
}\mathcal{Q})\\
&=&\sum_{\begin{array}{c} wt(jvt)+wt(-jv)+wt(j)\\
=2n+1, 0<j<3^n-1\end{array}}(-1)^{jv}\sigma(jvt)\sigma(-jv)\sigma(j)(\gamma\lambda^{vt})^j.
\end{eqnarray*}
Thus the assertion is established.
\done

\begin{remark}
Assume that $(v, t)$ is a realizable pair. By Theorem \ref{thm_general}, for $(\lambda, \gamma)\neq(\lambda_1, \gamma_1)$, if $\gamma\lambda^{vt}=\gamma_1\lambda_1^{vt}$, then $g(v,t)(\lambda, \gamma)=g(v,t)(\lambda_1, \gamma_1)$.
\end{remark}

With notations as in Theorem \ref{thm_general}, let $U=\{x^{vt}|x\in \mathbb{F}_{3^n}^{*}\}(=\{x^d|x\in \mathbb{F}_{3^n}^{*}\})$, and $\Lambda=\{\gamma_0, \gamma_1, \cdots  , \gamma_{d-1}\}$ be a set of representatives for the cosets of $U$ in $\mathbb{F}_{3^n}^{*}$, i.e., $\mathbb{F}_{3^n}^{*}=\gamma_0U\cup\gamma_1U\cup\cdots  \cup\gamma_{d-1}U$. Let $\alpha$ be a primitive element of $\mathbb{F}_{3^n}$. For any $0\leq i<3^n-1$, $\alpha^i$ can be written in the form of $\alpha^i=\gamma\lambda^{vt}$, where $\gamma\in \Lambda$ and $\lambda\in \mathbb{F}_{3^n}$. Then we can construct a ternary sequence $T=\{t_i\}$ by
\begin{equation}\label{eqn_T}
t_i=g(v,t)(\lambda, \gamma), i=0, 1, 2, \cdots
\end{equation}
Note that for any $(\lambda_1, \gamma_1)\neq(\lambda, \gamma)$, if $\gamma_1\lambda_1^{vt}=\gamma\lambda^{vt}$, then $t_i=g(v,t)(\lambda_1, \gamma_1)$.

\begin{theorem}\label{thm_T}
The ternary sequence $T=\{t_i\}$ defined by (\ref{eqn_T}) is an ideal two-level autocorrelation sequence.
\end{theorem}

{\bf Proof. }For any $0\leq i<3^n-1$ and $0<\tau<3^n-1$, let $\alpha^i=\gamma\lambda^{vt}$, and $\alpha^\tau=\widetilde{\gamma}\widetilde{\lambda}^{vt}$, where $\gamma, \widetilde{\gamma}\in \Lambda$, and $\lambda, \widetilde{\lambda}\in \mathbb{F}_{3^n}$. Then $\alpha^{i+\tau}=(\gamma\widetilde{\gamma})(\lambda\widetilde{\lambda})^{vt}$. Thus, $t_{i+\tau}=g(v,t)(\lambda\widetilde{\lambda}, \gamma\widetilde{\gamma})$. We have
\begin{eqnarray*}
C_T(\tau)&=&\sum_{i=0}^{3^n-2}\omega_3^{t_{i+\tau}-t_i}\\
&=&\frac{1}{d}\sum_{\gamma\in \Lambda}\sum_{\lambda\in \mathbb{F}_{3^n}^{*}}\omega_3^{g(v,t)(\lambda\widetilde{\lambda}, \gamma\widetilde{\gamma})-g(v,t)(\lambda, \gamma)}.
\end{eqnarray*}
According to the definition of $g(v,t)(\lambda, \gamma)$, we have
\begin{eqnarray*}
C_S(\tau)&=&\frac{1}{3^{2n}d}\sum_{\gamma\in \Lambda}\sum_{\lambda\in \mathbb{F}_{3^n}^{*}}\widehat{f}(v,t)(\lambda\widetilde{\lambda}, \gamma\widetilde{\gamma})\overline{\widehat{f}(v,t)(\lambda, \gamma)}\\
&=&\frac{1}{3^{2n}d}\sum_{\gamma\in \Lambda}\sum_{\lambda\in \mathbb{F}_{3^n}}\widehat{f}(v,t)(\lambda\widetilde{\lambda}, \gamma\widetilde{\gamma})\overline{\widehat{f}(v,t)(\lambda, \gamma)}-1\\
&=&\frac{1}{3^{2n}d}\sum_{\gamma\in \Lambda}\sum_{\lambda\in \mathbb{F}_{3^n}}\sum_{x_1, y_1\in \mathbb{F}_{3^n}}\omega_3^{Tr(\lambda\widetilde{\lambda} y_1)-Tr(y_1^tx_1)+Tr(\gamma\widetilde{\gamma} x_1^v)}\sum_{x_2, y_2\in \mathbb{F}_{3^n}}\omega_3^{-Tr(\lambda y_2)+Tr(y_2^tx_2)-Tr(\gamma x_2^v)}-1\\
&=&\frac{1}{3^nd}\sum_{\gamma\in \Lambda}\sum_{x_1, x_2, y\in \mathbb{F}_{3^n}}\omega_3^{-Tr(y^tx_1)+Tr(\gamma\widetilde{\gamma} x_1^v)+Tr(\widetilde{\lambda}^ty^tx_2)-Tr(\gamma x_2^v)}-1\\
&=&\frac{1}{d}\sum_{\gamma\in \Lambda}\sum_{x_2\in \mathbb{F}_{3^n}}\omega_3^{Tr(\gamma\widetilde{\gamma}\widetilde{\lambda}^{vt} x_2^v)-Tr(\gamma x_2^v)}-1\\
&=&\frac{1}{d}\sum_{\gamma\in \Lambda}\sum_{x_2\in \mathbb{F}_{3^n}}\omega_3^{Tr((\alpha^\tau-1)\gamma x_2^v)}-1\\
&=&\sum_{x\in \mathbb{F}_{3^n}}\omega_3^{Tr((\alpha^\tau-1)x)}-1=-1.
\end{eqnarray*}
\done

\begin{theorem}\label{thm_hamming}
For any $n=2m+1$, let $v=2(3^{m+1}-1)$, and $t=(3^n+1)/4$. Then $wt(jvt)+wt(-jv)+wt(j)>2n$ for any $0<j<3^n-1$. Moreover, $wt(jvt)+wt(-jv)+wt(j)=2n+1$ if and only if $j\in \{3^i, (2\cdot3^m+1)3^i\ |\ i=0, 1, \cdots  , n-1\}$.
\end{theorem}

The proof of Theorem \ref{thm_hamming} is heavily related to the enumerating  techniques for computing the Hamming weights of  ternary numbers $jvt, -jv$ and $j$. So we  postpone it  to  Section \ref{sec_lin_2}.

\begin{theorem}
The Lin conjecture is true.
\end{theorem}

{\bf Proof. }Let $n=2m+1$, $v=2(3^{m+1}-1)$, and $t=(3^n+1)/4$. Then $\gcd(v, 3^n-1)=2$ and $\gcd(t, 3^n-1)=1$.

Let $f(x)=Tr(x)$. By Theorem \ref{thm_general}, $(v, t)$ is a realizable pair, and
\begin{eqnarray*}
g(v,t)(\lambda, \gamma)&=&\sum_{0<j<3^n-1:\ wt(jvt)+wt(-jv)+wt(j)=2n+1}(-1)^{jv}\sigma(jvt)\sigma(-jv)\sigma(j)(\gamma\lambda^{vt})^j\\
&=&\sum_{0<j<3^n-1:\ wt(jvt)+wt(-jv)+wt(j)=2n+1}\sigma(jvt)\sigma(-jv)\sigma(j)(\gamma\lambda^{vt})^j.
\end{eqnarray*}
By Theorem \ref{thm_hamming},
\begin{eqnarray*}
g(v,t)(\lambda, \gamma)&=&\sum_{j\in \{3^i, (2\cdot3^m+1)3^i\ |\ i=0, 1, \cdots  , n-1\}}\sigma(jvt)\sigma(-jv)\sigma(j)(\gamma\lambda^{vt})^j\\
&=&2Tr(\gamma\lambda^{vt})+2Tr((\gamma\lambda^{vt})^{2\cdot3^m+1}).
\end{eqnarray*}
By Theorem \ref{thm_T}, $T=\{t_i\}$ constructed via (\ref{eqn_T}) has ideal two-level autocorrelation whose trace representation is given by
$$t_i=g(v,t)(\lambda, \gamma)=2Tr(\gamma\lambda^{vt})+2Tr((\gamma\lambda^{vt})^{2\cdot3^m+1})=2Tr(\alpha^i)+2Tr(\alpha^{(2\cdot3^m+1)i}).$$
We construct another $S=\{s_i\}$ where $s_i=2t_i$. Then $S$ also has ideal two-level autocorrelation. The trace representation of $S$ is given by
$$s_i=2t_i=Tr(\alpha^i)+Tr(\alpha^{(2\cdot3^m+1)i}).$$
Thus, the validity of the Lin  conjecture is established.\done

\begin{remark}{\em
The results in Theorems 2 and 3 are general, which state a relationship between the second order multiplexing DHT and ternary 2-level autocorrelation sequences with their trace representation.}
\end{remark}

\section{Proof of the Lin Conjecture: Part II}\label{sec_lin_2}

Theorem \ref{thm_hamming} is equivalent to the following theorem.

\begin{theorem}\label{thm_hamming_2}
Let $n=2m+1$. Then $wt(j)+wt((3^{m+1}-1)j)-wt(2(3^{m+1}-1)j)>0$ for any $0<j<3^n-1$. Moreover, $wt(j)+wt((3^{m+1}-1)j)-wt(2(3^{m+1}-1)j)=1$ if and only if $j\in \{3^i, (2\cdot3^m+1)3^i\ |\ i=0, 1, \cdots  , n-1\}$.
\end{theorem}

We need some preparations in order to prove this theorem. One may check that $wt(3j)=wt(j)$. We define $H(j)=wt(j)+wt((3^{m+1}-1)j)-wt(2(3^{m+1}-1)j)$.   We use $C_i$ to denote the coset modular $3^n-1$ which contains $i$. Thus $C_1=\{1, 3, \cdots, 3^{n-1}\}$ and $C_{2\cdot3^m+1}=\{ (2\cdot3^m+1)3^i \bmod{(3^n-1)}\, \,|\,\, i=0, 1, \cdots, n-1\}$,  since $\gcd(2\cdot3^m+1, 3^n-1)=1$.

For any $a>0$, we denote the residue of $a$ modulo $3^n-1$ by $\overline{a}$, i.e., $a\equiv \overline{a}\ (\bmod\;3^n-1)$ and $0\leq \overline{a}<3^n-1$. If $\overline{a}=\sum_{i=0}^{2m}a_{i}3^{i}$ with $a_{i}\in \{0, 1, 2\}$, then we write it as $\overline{a}=a_{2m}a_{2m-1}\cdots  a_1a_0$ for simplicity, i.e., $a_{2m}a_{2m-1}\cdots  a_1a_0$ is the ternary representation of $\overline{a}$.  However, if it is clear that $0<a<3^n-1$, sometimes, we also directly write $a$ instead of $\overline{a}$ for simplicity. For any $0\leq i\leq 2m$,  using the shift operation, we define an {\em equivalent relationship} on: $(a_{2m}a_{2m-1}\cdots  a_1a_0)\sim (a_ia_{i-1}\cdots  a_0a_{2m}\cdots  a_{i+1})$. The shift operation does not change the value of $H(j)$, i.e., we have
\begin{equation}\label{coset}
H(3^ij)=H(j), i=0, 1, \cdots, n-1, 0<j<3^n-1.
\end{equation}
Thus, for the assertion of Theorem \ref{thm_hamming_2},  we only need to show that for one $j$ in its equivalent class.

We need two more notations. For any $r\geq 0$, let
$$R_{r0}=\underbrace{11\cdots 11}_r0\mbox{ and }R_{r2}=\underbrace{11\cdots 11}_r2.$$
Then $\overline{a}\sim b_{t-1}b_{t-2}\cdots  b_0$, where $b_i=R_{r_i0}$ or $R_{r_i2}$, $i=0, 1, \cdots  , t-1$, and $t\geq 1$.

\begin{lemma}\label{lem_hamming_sum}
With notations as above, $wt(\overline{2a})=\sum_{i=0}^{t-1}wt(\overline{2b_i})$.
\end{lemma}

{\bf Proof. }

1) If $b_i=R_{r_i0}$  for all $0\leq i\leq t-1$, the result follows immediately.

2) If $b_i=R_{r_i2}$  for all $0\leq i\leq t-1$, then
$$\overline{2a}=\overline{2(\underbrace{11\cdots  1}_{r_{t-1}}2\underbrace{11\cdots  1}_{r_{t-2}}2\cdots  \underbrace{11\cdots  1}_{r_0}2)}=\underbrace{00\cdots  0}_{r_{t-1}}2\underbrace{00\cdots  0}_{r_{t-2}}2\cdots  \underbrace{00\cdots  0}_{r_0}2.$$
Hence, $wt(\overline{2a})=2t=\sum_{i=0}^{t-1}2=\sum_{i=0}^{t-1}wt(\overline{2b_i})$.

3) If these exist $0\leq i\neq j\leq t-1$ such that $b_i=R_{r_i0}$ and $b_j=R_{r_j2}$, we may assume that $b_{t-1}=R_{r_{t-1}0}$ and $b_0=R_{r_02}$. In this case, $\overline{2a}=2\overline{a}$. Let us compute
$$2\overline{a}=2(b_{t-1}b_{t-2}\cdots  b_0)=2(b_{t-1}b_{t-2}\cdots  b_1\underbrace{00\cdots  00}_{r_0+1})+2(\underbrace{11\cdots  1}_{r_0}2)=2(b_{t-1}b_{t-2}\cdots  b_1\underbrace{00\cdots  00}_{r_0+1})+1\underbrace{00\cdots  00}_{r_0}1.$$
Because the last digit of $2(b_{t-1}b_{t-2}\cdots  b_1)$ is 0 or 1, we get
$$wt(2\overline{a})=wt(2(b_{t-1}b_{t-2}\cdots  b_1))+2=wt(2(b_{t-1}b_{t-2}\cdots  b_1))+wt(2b_0).$$
Similarly, $wt(2(b_{t-1}b_{t-2}\cdots  b_1))=wt(2(b_{t-1}b_{t-2}\cdots  b_2))+wt(2b_1)$, and so on. Hence, $wt(2\overline{a})=\sum_{i=0}^{t-1}wt(2b_i).$

\done

This lemma shows that  the Hamming weight of $\overline{2a}$ can be computed through the Hamming weights of  their the runs of 1's. Here the runs of 1's play an important rule in  computing $H(j)$.

\begin{lemma}\label{lem_double_difference}

\begin{description}
\item[(i)] For any $r\geq 0$, $wt(R_{r0})-wt(2R_{r0})=-r$, and $wt(R_{r2})-wt(2R_{r2})=r$.
\item[(ii)]  If $R_{r0}$ and $R_{r2}$ appear as a pair in $\overline{a}$, then $wt(\overline{a})= wt(\overline{2a})$.
\end{description}
\end{lemma}

{\bf Proof. }The proof is easy, so we omit it.

\done

Note that we allow $r=0$.  Thus we have  $wt(R_{02})= wt(\overline{2R_{02}})$.   This lemma is another important counting technique for the Hamming weights of  $\overline{a}$ and $\overline{2a}$, which will be frequently used later. The following lemma shows that the effect of changing digits in $a$.

\begin{lemma}\label{lem_sum_2}
For $i\geq 0$, $wt(\overline{a+2\cdot3^i})-wt(\overline{2(a+2\cdot3^i)})\geq wt(\overline{a})-wt(\overline{2a})-2$.
\end{lemma}

{\bf Proof. }
Assume that $\overline{a}\sim b_{t-1}b_{t-2}\cdots  b_0$, where $b_i=R_{r_i0}$ or $R_{r_i2}$, $i=0, 1, \cdots  , t-1$, and $t\geq 1$. Under this equivalence, without loss of generality, we still keep the notation of $i$, and assume that $0\leq i<n$. In the following, if $j\geq t$, then $b_j=b_{j-t}$; if $j<0$, then $b_j=b_{j+t}$.

Let $\Delta=wt(\overline{a+2\cdot3^i})-wt(\overline{2(a+2\cdot3^i)})-[wt(\overline{a})-wt(\overline{2a})]$. Let us look at $\overline{a}+2\cdot3^i$, which is actually the addition of $2$ to one digit of some $b_k$, where $0\leq k\leq t-1$.
In the following, we consider $b_k=R_{r0}$ and $b_k=R_{r2}$ separately, since for each case,  the location of a digit which will be changed effects the Hamming weights of the resultant number.

Let $a_{i+v-1}, \cdots, a_{i+1}, a_i$ is a segment of $a$. We say that $a_i$ is the least significant digit (LSD) of the segment and $a_{i+v-1}$, the most significant digit (MSD) of the segment.

\paragraph{Case 1.}  $b_k=R_{r0}$.

(1) 2 is added to the LSD of  $b_k$: In this case, $b_k=\underbrace{11\cdots  11}_r0\rightarrow\underbrace{11\cdots  11}_r2$. By Lemmas \ref{lem_hamming_sum} and \ref{lem_double_difference}, $\Delta=r-(-r)=2r>-2$.

(2) 2 is added to the MSD of $b_k$.

i) $b_j=2$ for any $j\neq k$: In this case, $\overline{\overline{a}+2\cdot3^i}=00\cdots  0\underbrace{11\cdots  1}_r00\cdots  0$. By Lemmas \ref{lem_hamming_sum} and \ref{lem_double_difference}, $\Delta=-r-(-r)=0>-2$.

ii) $b_{k+1}=b_{k+2}=\cdots  =b_{k+j}=2, b_{k+j+1}=R_{p0}$: $\underbrace{11\cdots  11}_p0\underbrace{22\cdots  22}_j\underbrace{11\cdots  11}_r0\rightarrow \underbrace{11\cdots  11}_{p+1}\underbrace{00\cdots  00}_{j+1}\underbrace{11\cdots  11}_{r-1}0$. By Lemmas \ref{lem_hamming_sum} and \ref{lem_double_difference}, $\Delta=-(p+1)+(-(r-1))-(-p+(-r))=0>-2$.

iii) $b_{k+1}=b_{k+2}=\cdots  =b_{k+j}=2, b_{k+j+1}=R_{p2}$: $\underbrace{11\cdots  11}_p0\underbrace{22\cdots  22}_j\underbrace{11\cdots  11}_r2\rightarrow \underbrace{11\cdots  11}_{p+1}\underbrace{00\cdots  00}_{j+1}\underbrace{11\cdots  11}_{r-1}2$. By Lemmas \ref{lem_hamming_sum} and \ref{lem_double_difference}, $\Delta=-(p+1)+(r-1)-(-p+r)=-2$.

(3) 2 is added to one middle digit of $b_k$: $b_k=\underbrace{11\cdots  11}_r0\rightarrow\underbrace{11\cdots  1}_{r_1}20\underbrace{11\cdots  1}_{r_2}0$, where $r_1+r_2=r-2$. By Lemmas \ref{lem_hamming_sum} and \ref{lem_double_difference}, $\Delta=r_1-r_2-(-r)=2r_1+2>-2$.

\paragraph{Case 2.}  $b_k=R_{r2}$.

(1) 2 is added to the LSD of $b_k$.

i) $r>0$ and $b_{k-1}=R_{t0}$: $\underbrace{11\cdots  11}_r2\underbrace{11\cdots  111}_t0\rightarrow\underbrace{11\cdots  11}_{r-1}21\underbrace{11\cdots  111}_t0$. By Lemmas \ref{lem_hamming_sum} and \ref{lem_double_difference}, $\Delta=(r-1)+(-(t+1))-(r+(-t))=-2$.

ii) $r>0$ and $b_{k-1}=R_{t2}$: $\underbrace{11\cdots  11}_r2\underbrace{11\cdots  111}_t2\rightarrow\underbrace{11\cdots  11}_{r-1}21\underbrace{11\cdots  111}_t2$. By Lemmas \ref{lem_hamming_sum} and \ref{lem_double_difference}, $\Delta=(r-1)+(t+1)-(r+t)=0$.

iii) $b_k=b_{k+1}=\cdots  =b_{k+j}=2, b_{k+j+1}=R_{p2}, b_{k-1}=R_{t0}$: $\underbrace{11\cdots  11}_p\underbrace{22..2}_{j+2}\underbrace{11\cdots  1}_t0\rightarrow\underbrace{11\cdots  11}_{p-1}2\underbrace{00..0}_{j+1}\underbrace{11\cdots  1}_{t+1}0$. By Lemmas \ref{lem_hamming_sum} and \ref{lem_double_difference}, $\Delta=(p-1)+(-(t+1))-(p+(-t))=-2$.

iv) $b_k=b_{k+1}=\cdots  =b_{k+j}=2, b_{k+j+1}=R_{p2}, b_{k-1}=R_{t2}$: $\underbrace{11\cdots  11}_p\underbrace{22..2}_{j+2}\underbrace{11\cdots  1}_t2\rightarrow\underbrace{11\cdots  11}_{p-1}2\underbrace{00..0}_{j+1}\underbrace{11\cdots  1}_{t+1}2$. By Lemmas \ref{lem_hamming_sum} and \ref{lem_double_difference}, $\Delta=(p-1)+(t+1)-(p+t)=0$.

v) $b_k=b_{k+1}=\cdots  =b_{k+j}=2, b_{k+j+1}=R_{p0}, b_{k-1}=R_{t0}$: $\underbrace{11\cdots  11}_p0\underbrace{22..2}_{j+1}\underbrace{11\cdots  1}_t0\rightarrow\underbrace{11\cdots  11}_{p+1}\underbrace{00..0}_{j}\underbrace{11\cdots  1}_{t+1}0$. By Lemmas \ref{lem_hamming_sum} and \ref{lem_double_difference}, $\Delta=-(p+1)+(-(t+1))-(-p-t)=-2$.

vi) $b_k=b_{k+1}=\cdots  =b_{k+j}=2, b_{k+j+1}=R_{p0}, b_{k-1}=R_{t2}$: $\underbrace{11\cdots  11}_p0\underbrace{22..2}_{j+1}\underbrace{11\cdots  1}_t2\rightarrow\underbrace{11\cdots  11}_{p+1}\underbrace{00..0}_{j}\underbrace{11\cdots  1}_{t+1}2$. By Lemmas \ref{lem_hamming_sum} and \ref{lem_double_difference}, $\Delta=-(p+1)+(t+1)-(-p+t)=0$.

(2) 2 is added to the MSD of $b_k$.

i) $b_j=2$ for any $j\neq k$: In this case, $\overline{\overline{a}+2\cdot3^i}=00\cdots  0\underbrace{11\cdots  1}_{r-2}20\cdots  0$, where $r\geq 2$, or $00\cdots  0100\cdots  0$, where $r=1$. Thus, $\Delta=r-2-r=-2$ or $\Delta=-1-1=-2$.

ii) $b_{k+1}=b_{k+2}=\cdots  =b_{k+j}=2, b_{k+j+1}=R_{p0}$: $\underbrace{11\cdots  11}_p0\underbrace{22\cdots  2}_{j}\underbrace{11\cdots  1}_r2\rightarrow\underbrace{11\cdots  11}_{p+1}\underbrace{00..0}_{j+1}\underbrace{11\cdots  1}_{r-1}2$. By Lemmas \ref{lem_hamming_sum} and \ref{lem_double_difference}, $\Delta=-(p+1)+(r-1)-(-p+r)=-2$.

iii) $b_{k+1}=b_{k+2}=\cdots  =b_{k+j}=2, b_{k+j+1}=R_{p2}$: $\underbrace{11\cdots  11}_p\underbrace{22\cdots  2}_{j+1}\underbrace{11\cdots  1}_r2\rightarrow\underbrace{11\cdots  11}_{p-1}2\underbrace{00..0}_{j+2}\underbrace{11\cdots  1}_{r-1}2$. By Lemmas \ref{lem_hamming_sum} and \ref{lem_double_difference}, $\Delta=(p-1)+(r-1)-(p+r)=-2$.

(3) 2 is added to the middle digit of $b_k$: $b_k=\underbrace{11\cdots  11}_r2\rightarrow\underbrace{11\cdots  1}_{r_1}20\underbrace{11\cdots  1}_{r_2}2$, where $r_1+r_2=r-2$. By Lemmas \ref{lem_hamming_sum} and \ref{lem_double_difference}, $\Delta=r_1-r_2-r=-2$.

\done

\begin{lemma}\label{lem_simple}
For any $i, j>0$, we have $$(3^{m+1}-1)[j\pm(3^{m+1}+1)3^i]\equiv (3^{m+1}-1)j\pm2\cdot3^i\ (\bmod\;3^n-1).$$
\end{lemma}

{\bf Proof. }
\begin{eqnarray*}
(3^{m+1}-1)[j\pm(3^{m+1}+1)3^i]&\equiv& (3^{m+1}-1)j\pm(3^{2m+2}-1)3^i\ (\bmod\;3^n-1)\\
&\equiv&(3^{m+1}-1)j\pm2\cdot3^i\ (\bmod\;3^n-1).
\end{eqnarray*}\done

Let $\overline{a}\sim b_{t-1}b_{t-2}\cdots  b_0$, where $b_i=R_{r_i0}$ or $R_{r_i2}$, $i=0, 1, \cdots  , t-1$, and $t\geq 1$. In the following lemma,  we present the result on    how  $wt(\overline{2a})$ will  be changed when one digit is changed in $\overline{a}$.

\begin{lemma}\label{lem_onedigit_difference}
Suppose that the segment to be changed is $b_i$. We denote the resulting ternary vector from $\overline{a}$ by $\overline{a^{'}}$.
\begin{enumerate}
  \item[1)] $b_i=R_{r0}$, $b_i^{'}=2\underbrace{11\cdots  11}_{r-1}0$: $\Delta=wt(\overline{2a^{'}})-wt(\overline{2a})=0$;
  \item[2)] $b_i=R_{r2}$, $b_i^{'}=\underbrace{11\cdots  11}_{r_1}0\underbrace{11\cdots  11}_{r_2}2$ where $r_1+r_2=r-1$: $\Delta=wt(\overline{2a^{'}})-wt(\overline{2a})=2r_1$;
  \item[3)] $b_i=R_{r2}$, $b_i^{'}=2\underbrace{11\cdots  11}_{r-1}2$: $\Delta=wt(\overline{2a^{'}})-wt(\overline{2a})=2$;
  \item[4)] $b_i=R_{r2}, b_{i-1}=0$, $b_i^{'}=\underbrace{11\cdots  11}_{r+1}$: $\Delta=wt(\overline{2a^{'}})-wt(\overline{2a})=2r$.
\end{enumerate}
\end{lemma}

{\bf Proof. }By Lemma \ref{lem_double_difference}, the result follows immediately.\done

\bigskip

Now we are ready to show a proof of  Theorem \ref{thm_hamming_2}.

{\bf Proof of Theorem \ref{thm_hamming_2}. } The proof consists of two parts. First we show that $H(j)=1$ when $J\in C_1\cup C_{2\cdot 3^m+1}$. Then we show that $H(j)\ge 2$ if $j\notin C_1\cup C_{2\cdot 3^m+1}$. In other words, we have the following statements.  \vspace{0.1in}

{\bf Claim 1.}  If $j \in C_1\cup C_{2\cdot 3^m+1}$, then $H(j)=1$.    This can be easily verified. \vspace{0.1in}

{\bf Claim 2.}  If   $j\notin C_1\cup C_{2\cdot 3^m+1}$, then $H(j)\ge 2$.

\paragraph{\bf Proof of Claim  2.}
We will use the induction to show this result.  Note that for $j=2$, we have $H(j)=2$. Assume that Claim 2  holds for $2\leq j\leq k-1<3^n-1$. Now we consider the case of $j=k$. We write  $j=a_{2m}a_{2m-1}\cdots  a_1a_0$, the ternary representation of $j$. In the following, if $i>2m$, then $a_{i}=a_{i-2m-1}$.
In order to compute $wt(\overline{(3^{m+1}-1)j})$  and  $wt(\overline{2(3^{m+1}-1)j})$, we  need to consider $2m+1$ pairs: $(a_0, a_{m+1}), \cdots  , (a_i, a_{m+1+i}), \cdots, (a_{2m}, a_{m})$ from counting the Hamming weight of $(3^{m+1}-1)j$,  i.e.,
\[
\begin{array}{||c|cccc|c|cccc||}
\hline
3^{m+1}j& a_{m-1}& a_{m-2}& \cdots & a_0 & a_{2m} & a_{2m-1} & \cdots & a_{m+1} & a_m\\
j&a_{2m} & a_{2m-1} & \cdots & a_{m+1} & a_m & a_{m-1}&  \cdots & a_1 &a_0 \\ \hline
\end{array}
\]

\paragraph{Type 1.} There exists $0\leq i \leq 2m$ such that $a_{i }\neq 0$ and $a_{m+1+i}\neq 0$.

If $j=(3^n-1)/2$, then $H(j)>1$. Hence, we can assume that $j\neq(3^n-1)/2$.
Let $j'=j-\overline{(3^{m+1}+1)3^{i}}$. Then $0\leq j'<j$ and $wt(j)=wt(j')+2$. If $j'=0$, then $H(j)=2$. Otherwise, by Lemmas \ref{lem_sum_2} and \ref{lem_simple}, we have the following inequalities:
\begin{eqnarray*}
H(j)&=&H(j'+\overline{(3^{m+1}+1)3^{i}})\\
&=&wt(\overline{(3^{m+1}-1)j'+2\cdot3^{i}})-wt(\overline{2((3^{m+1}-1)j'+2\cdot3^{i})})+wt(j')+2\\
&\geq&wt(\overline{(3^{m+1}-1)j'})-wt(\overline{2(3^{m+1}-1)j'})-2+wt(j')+2 \\
&= & H(j').
\end{eqnarray*}
Since $j'<j$, if $j'\notin C_1\cup C_{2\cdot 3^m+1}$, then $H(j')\ge 2$.  We now compute $H(j)$ directly for the case that  $j'\in C_1\cup C_{2\cdot 3^m+1}$. We only need to compute $j'=1$ and $j'=2\cdot 3^m+1$.

For $j'=1$, then $j=1+\overline{(3^{m+1}+1)3^{i}}\Longrightarrow wt(j) = 3$. If $i=0$, then $1+\overline{(3^{m+1}+1)3^{i}}=3^{m+1}+2\in C_{2\cdot 3^m+1}$. Hence, we may assume $i\neq 0$. In this case,
\begin{eqnarray*}
H(j)&=&wt(j)+wt((3^{m+1}-1)j)-wt(2(3^{m+1}-1)j)\\
&=&wt(1+(3^{m+1}+1)3^{i})+wt((3^{m+1}-1)(1+(3^{m+1}+1)3^{i}))\\
&&-wt(2(3^{m+1}-1)(1+(3^{m+1}+1)3^{i}))\\
&=&wt(1+(3^{m+1}+1)3^{i})+wt(3^{m+1}-1+2\cdot 3^{i}))-wt(2(3^{m+1}-1)+4\cdot 3^{i}))\\
&=&3+wt(3^{m+1}-1+2\cdot 3^{i}))-4\\
&=&wt(3^{m+1}-1+2\cdot 3^{i}))-1\\
&=&2i+2-1\\
&\geq&3.
\end{eqnarray*}
Similarly,  if $j'=2\cdot 3^m+1$, then $j=2\cdot 3^m+1+\overline{(3^{m+1}+1)3^{i}}\Longrightarrow wt(j) \ge 3$. If $i=m$, then $2\cdot 3^m+1+\overline{(3^{m+1}+1)3^{i}}=3^{m+1}+2\in C_{2\cdot 3^m+1}$. Hence, we may assume $i\neq m$. In this case, we also have
\begin{eqnarray*}
H(j)&=&wt(j)+wt((3^{m+1}-1)j)-wt(2(3^{m+1}-1)j)\\
&=&wt(2\cdot 3^m+1+(3^{m+1}+1)3^{i})+wt((3^{m+1}-1)(2\cdot 3^m+1+(3^{m+1}+1)3^{i}))\\
&&-wt(2(3^{m+1}-1)(2\cdot 3^m+1+(3^{m+1}+1)3^{i}))\\
&=&wt(2\cdot 3^m+1+(3^{m+1}+1)3^{i})+wt((3^{m+1}-1)(2\cdot 3^m+1)+2\cdot3^{i})\\
&&-wt(2(3^{m+1}-1)(2\cdot 3^m+1)+4\cdot3^{i})\\
&=&wt(2\cdot 3^m+1+(3^{m+1}+1)3^{i})+wt(3^m+1+2\cdot3^{i}))-wt(2(3^m+1)+4\cdot3^{i}))\\
&=&\left\{\begin{array}{ll}
     5+2-4=3, & i=0; \\
     5+4-4=5 & i=m-1; \\
     3+4-4=3 & i=2m; \\
     5+4-6=3 & i\neq 0, m-1, m, 2m.
   \end{array}\right.\\
&\geq&3.
\end{eqnarray*}

Thus Claim  2 is true for this case.

\paragraph{Type 2.}  For any $0\leq i\leq 2m$, $a_i=0$ or $a_{m+1+i}=0$.
Suppose that $a_i\leq a_{m+1+i}$ for $0\leq i\leq 2m$. Then, it follows that $a_0\leq a_{m+1}\leq a_1$, i.e., $a_0\leq a_1$. Similarly, we have $a_1\leq a_2\leq \cdots  \leq a_{2m}\leq a_0$. Thus, $a_{2m}=a_{2m-1}=\cdots=a_{0}$ which means that $j=(3^n-1)/2$. We get a contradiction.  Thus there exists $0\leq i_1\leq 2m$ such that $a_{i_1}>a_{m+1+i_1}$. As a consequence,
$$\overline{3^{2m-i_1}j}=a_{i_1}a_{i_1-1}\cdots a_0a_{2m}\cdots a_{i_1+1}>a_{m+i_1+1}a_{m+i_1}\cdots a_0a_{2m}\cdots a_{m+i_1+2}=\overline{3^{m-i_1-1}j}.$$
Because $H(j)=H(3j)$, without loss of generality, we can assume that $$\overline{3^{m+1}j}=a_{m-1}a_{m-2}\cdots a_1a_{0}\cdots a_{m}>a_{2m}a_{2m-1}\cdots a_1a_0=j.$$
In this case, $\overline{(3^{m+1}-1)j}=(a_{m-1}a_{m-2}\cdots a_1a_{0}\cdots a_{m})-(a_{2m}a_{2m-1}\cdots a_1a_0)$.   We can classify the ternary representation of  those $j$ into  three disjoint cases, which are listed
 in Table \ref{tab_pattern}.

\begin{table}[ht!]
\centering
\caption{Three Disjoint Cases of Patterns in the Ternary Representation of $j$ with $a_i=0$ or $a_{i+m-1}=0$ for all $0\le i\le  2m$}\label{tab_pattern}
\begin{tabular}{||c|c||}
 \hline
  &  Patterns \\
\hline
\hline
 Case I & $x\underbrace{aa\cdots aa}_{r\geq 2}0$: $a\neq0, x\neq a$ \\
\hline
 Case II & $x\underbrace{aa\cdots aa}_{r\geq 1}b0$: $a\neq 0, b\neq 0, a\neq b, x\neq a$ \\
\hline
 Case III & $0a0$: $a\neq0$ \\
\hline
\end{tabular}
\end{table}

{\bf Case I: }$j$ contains a segment of the form $x\underbrace{aa\cdots  aa}_{r\geq 2}0$, where $a\neq0, x\neq a$.

(1) $a=1$. In this case, $\overline{(3^{m+1}-1)j}$ contains two segments
$$
\begin{array}{ccc}
  \begin{array}{cccccc}
  &1& 1& \cdots  & 1& 0 \\
  -& 0& 0& \cdots  & 0& 0 \\
  \hline
  &d_1&d_2&\cdots  &d_r&d_{r+1}\\
  &(1&1&\cdots  &1&0)\\
  &(1&1&\cdots  &0&2)
\end{array} & \mbox{and} &
\begin{array}{ccccccc}
  &0& 0& 0& \cdots  & 0& y \\
  -& x& 1& 1& \cdots  & 1& 0 \\
  \hline
  &e_0&e_1&e_2&\cdots  &e_{r}&e_{r+1}\\
  &(e_0&1&1&\cdots  &2)&\\
  &(e_0&1&1&\cdots  &1&2)
\end{array}
\end{array}.
$$
In other words, $d_1=d_2=\cdots  =d_{r-1}=1$, $d_r=1$, $d_{r+1}=0$, or $d_r=0$, $d_{r+1}=2$; $e_1=e_2=\cdots  =e_{r-1}=1$, $e_r=2$, or $e_r=1, e_{r+1}=2$. Because $x\neq 1$, $e_0\neq 1$. We change the segment of $j$ from $\underbrace{11\cdots  11}_{r\geq 2}0$ to $\underbrace{01\cdots  11}_{r\geq 2}0$, and denote the new integer by $j^{'}$. Then $d_1^{'}=0, e_1^{'}=2$, and other $d_i, e_i$ stay the same. Therefore, $wt((3^{m+1}-1)j^{'})=wt((3^{m+1}-1)j)$. By Lemma \ref{lem_onedigit_difference}, $wt(2(3^{m+1}-1)j^{'})\geq wt(2(3^{m+1}-1)j)$. Moreover, $wt(j)=wt(j^{'})+1$. Therefore, $H(j)\geq H(j^{'})+1\geq 2$.

(2) $a=2$. In this case, $\overline{(3^{m+1}-1)j}$ contains two segments
$$
\begin{array}{ccc}
  \begin{array}{cccccc}
  &2& 2& \cdots  & 2& 0 \\
  -& 0& 0& \cdots  & 0& 0 \\
  \hline
  &d_1&d_2&\cdots  &d_r&d_{r+1}\\
  &(2&2&\cdots  &1&2)\\
  &(2&2&\cdots  &2&0)
\end{array} & \mbox{and} &
\begin{array}{ccccccc}
  &0& 0& 0& \cdots  & 0& y \\
  -& x& 2& 2& \cdots  & 2& 0 \\
  \hline
  &e_0&e_1&e_2&\cdots  &e_{r}&e_{r+1}\\
  &(e_0&0&0&\cdots  &0&2)\\
  &(e_0&0&0&\cdots  &1&y)
\end{array}
\end{array}.
$$
In other words, $d_1=d_2=\cdots  =d_{r-1}=2$, $d_r=1$, $d_{r+1}=2$, or $d_r=2$, $d_{r+1}=0$; $e_0=1$ or 2, $e_1=e_2=\cdots  =e_{r-1}=0$, $e_r=0$, $e_{r+1}=2$, or $e_r=1$, $e_{r+1}=y$ or $y-1$. We change the segment from $0\underbrace{22\cdots  22}_{r\geq 2}0$ to $0\underbrace{22\cdots  21}_{r\geq 2}0$, and denote the new integer by $j^{'}$. Then $d_r^{'}=d_r-1, e_r^{'}=e_r+1$, and other $d_i, e_i$ stay the same. Therefore, $wt((3^{m+1}-1)j^{'})=wt((3^{m+1}-1)j)$. By Lemma \ref{lem_onedigit_difference}, $wt(2(3^{m+1}-1)j^{'})\geq wt(2(3^{m+1}-1)j)$. Moreover, $wt(j)=wt(j^{'})+1$. Therefore, $H(j)\geq H(j^{'})+1\geq 2$.

{\bf Case II: }$j$ contains a segment of the form $x\underbrace{aa\cdots  aa}_{r\geq 1}b0$, where $a\neq 0, b\neq 0, a\neq b, x\neq a$.

(1) $a=1, b=2$.  Similarly, $\overline{(3^{m+1}-1)j}$ contains two segments
$$
\begin{array}{ccc}
  \begin{array}{ccccccc}
  &1& 1& \cdots  & 1& 2& y \\
  -& 0& 0& \cdots  & 0& 0& 0 \\
  \hline
  &d_1&d_2&\cdots  &d_r&d_{r+1}&d_{r+2}\\
   &(1&1&\cdots  &1&2)&\\
    &(1&1&\cdots  &1&1&2)
\end{array} & \mbox{and} &
\begin{array}{ccccccc}
  &0& 0& 0& \cdots  & 0& 0 \\
  -& x& 1& 1& \cdots  & 1& 2 \\
  \hline
  &e_0&e_1&e_2&\cdots  &e_{r}&e_{r+1}\\
   &(e_0&1&1&\cdots  &1&e_{r+1})
\end{array}
\end{array}.
$$
In other words, $d_1=d_2=\cdots  =d_r=1$, $d_{r+1}=1$ or $2$. If $d_{r+1}=1$, then $d_{r+2}=2$. Hence, $d_1d_2\cdots  d_{r+1}$ or $d_1d_2\cdots  d_{r+1}d_{r+2}$ is contained in a segment of form $R_{r2}$. $e_1=e_2=\cdots  =e_r=1$, $e_{r+1}=0$ or 1.
Because $x\neq 1$, $e_0\neq 1$. We change the segment of $j$ from $\underbrace{11\cdots  11}_{r\geq 1}2$ to $\underbrace{01\cdots  11}_{r\geq 1}2$, and denote the new integer by $j^{'}$. Then $d_1^{'}=0, e_1^{'}=2$, and the  other $d_i$'s and  $e_i$'s remain unchanged. Therefore, $wt((3^{m+1}-1)j^{'})=wt((3^{m+1}-1)j)$. By Lemma \ref{lem_onedigit_difference}, $wt(2(3^{m+1}-1)j^{'})\geq wt(2(3^{m+1}-1)j)$. Moreover, $wt(j)=wt(j^{'})+1$. Therefore, $H(j)\geq H(j^{'})+1\geq 2$.

(2) $a=2, b=1$. By the analysis above, we only need to consider the case of $x=0$. In this case, $\overline{(3^{m+1}-1)j}$ contains two segments
$$
\begin{array}{ccc}
  \begin{array}{ccccccc}
  &2& 2& \cdots  & 2& 1& 0 \\
  -& 0& 0& \cdots  & 0& 0& 0 \\
  \hline
  &d_1&d_2&\cdots  &d_r&d_{r+1}&d_{r+2}\\
  &(2&2&\cdots  &2&1&0)\\
  &(2&2&\cdots  &2&0&2)
\end{array} & \mbox{and} &
\begin{array}{cccccccc}
  &0& 0& 0& \cdots  & 0& 0& y \\
  -& 0& 2& 2& \cdots  & 2& 1& 0 \\
  \hline
  &e_0&e_1&e_2&\cdots  &e_r&e_{r+1}&e_{r+2}\\
   &(2&0&0&\cdots  &0&2)&\\
   &(2&0&0&\cdots  &0&1&2)
\end{array}
\end{array}.
$$

In other words, $d_1=d_2=\cdots  =d_{r}=2$, $d_{r+1}=1$, $d_{r+2}=0$, or $d_{r+1}=0$, $d_{r+2}=2$; $e_0=2$, $e_1=e_2=\cdots  =e_{r}=0$, $e_{r+1}=2$, or $e_{r+1}=1$, $e_{r+2}=2$. We change the segment of $j$ from $0\underbrace{22\cdots  22}_{r\geq 1}10$ to $0\underbrace{12\cdots  22}_{r\geq 1}10$, and denote the new integer by $j^{'}$. Then $d_1^{'}=1, e_1^{'}=1$, and the other $d_i$'s and $e_i$'s are unchanged. Therefore, $wt((3^{m+1}-1)j^{'})=wt((3^{m+1}-1)j)$. By Lemma \ref{lem_onedigit_difference}, $wt(2(3^{m+1}-1)j^{'})\geq wt(2(3^{m+1}-1)j)$. Moreover, $wt(j)=wt(j^{'})+1$. Therefore, $H(j)\geq H(j^{'})+1\geq 2$.

{\bf Case III: }$j$ contains $0$ and segments of the form $0a0$, where $a\neq0$.

(1) $j$ only contains $0$'s and segments of the form $010$. Since $j\not\in C_1$,   there are at least two segments of $010$. By Lemma \ref{lem_double_difference}, we only need to consider segments of form $S_{r0}$ in $\overline{(3^{m+1}-1)j}$. Among such patterns, one may check that only $S_{10}$ can occur:
$$  \begin{array}{r}
  1\ 0\ \cdots  \ 0\ 1\ 0 \\
  -\ 0\ 0\ \cdots  \ 0\ 0\ 0 \\
  \hline
  1\ 0\ \cdots  \ 0\ ?\ ?
\end{array}
$$
However, another segment also occurs:
$$  \begin{array}{r}
  0\ 0\ \cdots  \ 0\ 0 \\
  -\ 1\ 0\ \cdots  \ 0\ 1 \\
  \hline
  1\ 2\ \cdots  \ 2\ ?
\end{array}
$$
Therefore, ``10" and ``12" occur as a pair. Consequently, by Lemmas \ref{lem_hamming_sum} and \ref{lem_double_difference}, $wt((3^{m+1}-1)j)-wt(2(3^{m+1}-1)j)\geq 0$, and $H(j)\geq 2$.

(2) $j$ only contains $0$'s and segments of the form $020$.  Since $H(j)=2$ when $j=2$,  then  there are at least two segments of $020$. One may check that only $S_{110}$ and $S_{10}$ can occur. There are two cases.

i)
$$  \begin{array}{r}
  0\ 2\ 0\ 0 \\
  -\ ?\ 0\ 2\ 0 \\
  \hline
  ?\ 1\ 1\ 0
\end{array}
$$
However, this means that $(a_i, a_{i+m+1})=(2, 2)$ for certain $0\leq i\leq 2m$, which is impossible. Hence, $wt((3^{m+1}-1)j)-wt(2(3^{m+1}-1)j)\geq 0$, and $H(j)\geq 2$.

ii)
$$  \begin{array}{r}
  2\ 0\ \cdots  \ 0\ 0\ 0\ \cdots  \ 0\ 2\\
  -\ 0\ 0\ \cdots  \ 0\ 2\ 0\ \cdots  \ 0\ 0\\
  \hline
  1\ 2\ \cdots  \ 2\ 1\ 0\ \cdots  \ 0\ ?
\end{array}
$$
In this case, ``10" and ``12" occur as a pair. Consequently, by Lemmas \ref{lem_hamming_sum} and \ref{lem_double_difference}, $wt((3^{m+1}-1)j)-wt(2(3^{m+1}-1)j)\geq 0$, and $H(j)\geq 2$.

(3) $j$ contains $0$'s, and segments of both forms $020$ and $010$. There are 3 cases we need to consider.

i)
$$  \begin{array}{r}
  1\ 0\ \cdots  \ 0\ x\ 0 \\
  -\ 0\ 0\ \cdots  \ 0\ 0\ 0 \\
  \hline
  1\ 0\ \cdots  \ 0\ ?\ ?
\end{array}
$$ where $x=$ 1 or 2. In this case, by (1) of Case III, ``10" and ``12" occur as a pair. Consequently, by Lemmas \ref{lem_hamming_sum} and \ref{lem_double_difference}, $wt((3^{m+1}-1)j)-wt(2(3^{m+1}-1)j)\geq 0$, and $H(j)\geq 2$.

ii)

$$
\begin{array}{ccc}
\begin{array}{r}
  0\ 1\ 0\ 0 \\
  -\ ?\ 0\ 2\ 0 \\
  \hline
  ?\ 0\ 1\ 0
\end{array} & \mbox{or} &
\begin{array}{r}
  0\ 2\ 0\ 0 \\
  -\ ?\ 0\ 2\ 0 \\
  \hline
  ?\ 1\ 1\ 0
\end{array}
\end{array}.
$$
By (2) of Case III, this case is impossible.

iii)
$$  \begin{array}{r}
  x\ 0\ \cdots  \ 0\ 0\ 0\ \cdots  \ 0\ y\\
  -\ 0\ 0\ \cdots  \ 0\ 2\ 0\ \cdots  \ 0\ 0\\
  \hline
  ?\ 2\ \cdots  \ 2\ 1\ 0\ \cdots  \ 0\ ?
\end{array}
$$where $x=$ 1 or 2, $y=$ 1 or 2. If $x=1$, by (1) of Case III, ``10" and ``12" occur as a pair; if $x=2$, by (2) of Case III, ``10" and ``12" occur as a pair. Consequently, by Lemmas \ref{lem_hamming_sum} and \ref{lem_double_difference}, $wt((3^{m+1}-1)j)-wt(2(3^{m+1}-1)j)\geq 0$, and $H(j)\geq 2$.

\bigskip

According to Claims 1 and 2, the assertions of Theorem 6 is established.

\done

From Theorems 2-4,  the validity of Conjecture 2 in \cite{Gong2012} selected from \cite{five11} follows immediately.
\begin{corollary}
The Lin conjectured sequences are Hadamard equivalent to $m$-sequences.
\end{corollary}

\section{Concluding Remarks}\label{sec_con}

In this paper, we present a proof for the Lin conjecture  using the second order multiplexing DHT together with   Stickelberger's theorem,  and the Teichm\"{u}ller character for getting a sufficient and necessary condition for ideal  2-level autocorrelation sequences and their trace representation, and  combinatorial techniques for enumerating the Hamming weights of ternary numbers.   As we can see the treatments of the proof,  the results obtained in first part of the proof  is general, and the second part  of the proof is rather involved in  enumeration of the Hamming weights of ternary numbers.   As a by-product, we also confirmed a conjecture in \cite{five11}, which  is restated as Conjecture 2 in \cite{Gong2012}, i.e.,  two term sequences, conjectured by Lin, are Hadamard equivalent to $m$-sequences.
Furthermore, using the  second order multiplexing DHT, we have found  the realizable pairs of $(v, t)$ from starting an $m$-sequence instead of starting with a Lin sequence,  which realize the conjectured ideal two-level autocorrelation sequences in \cite{LG01} by computer search.  These new findings are under further investigation.

\section*{Acknowledgement} The third author wishes to thank  John Dillon for sending her their initial draft \cite{ArasuDillonPlayer} in November 2006.  All the authors of this paper would like   to thank Fei Huo and  Yang Yang for their participations of the  Waterloo Working Group for Attempting the Lin Conjecture  in August, 2011, Waterloo \cite{five11},  and their tremendous contributions and help for many computational results toward the proof.


\end{document}